\documentclass[aps,prl,twocolumn,amsmath,showpacs]{revtex4}
\usepackage{graphics,epsfig}
\usepackage{bm}

\begin{document}

\title{Orientation selection in lamellar phases by oscillatory shears} 
\author{Zhi-Feng Huang and Jorge Vi\~nals}
\affiliation{McGill Institute for Advanced Materials and Department of
Physics, McGill University, Montreal, QC H3A 2T8, Canada}

\date{\today}

\begin{abstract}
In order to address the selection mechanism that is responsible for the
unique lamellar orientation observed in block copolymers under oscillatory
shears, we use a constitutive law for the dissipative part of the stress
tensor that respects the uniaxial symmetry of a lamellar phase. An interface
separating two domains oriented parallel and perpendicular to the shear is
shown to be hydrodynamically unstable, a situation analogous to the
thin layer instability of stratified fluids under shear. The resulting
secondary flows break the degeneracy between parallel and perpendicular
lamellar orientation, leading to a preferred perpendicular
orientation in certain ranges of parameters of the polymer and of the
shear.

\end{abstract}

\maketitle

\begin{figure}
\includegraphics[width=2.in]{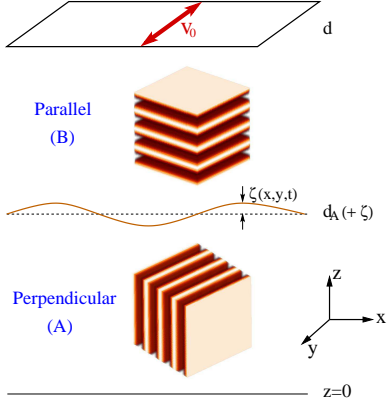}
\caption{Schematic representation of a parallel/perpendicular configuration
  begin sheared along $\hat{y}$.}
\label{fig_conf}
\end{figure}

Oscillatory shears are often used to promote long range order in lamellar 
phases of block copolymers, yet the mechanisms responsible for selecting a
particular lamellar
orientation relative to the shear remain unknown. A possible mechanism based
on a hydrodynamic instability in a microphase separated copolymer is presented
here that can distinguish between the experimentally important cases of
parallel and perpendicular orientations (Fig. \ref{fig_conf}). The instability
occurs at the boundary 
separating parallel and perpendicular regions provided that the dissipative
part of the stress tensor of the copolymer is chosen to reflect the uniaxial
symmetry 
of these broken symmetry phases. Our results rely solely on the uniaxial
symmetry of the microphases, and therefore should generally apply to other
complex fluids of the same symmetry.

Block copolymers are being extensively investigated as nanoscale templates for
a wide variety of applications that include nanolithography
\cite{re:harrison00b,re:black04,re:black05}, photonic components
\cite{re:urbas99}, or high density storage systems
\cite{re:thurn-albrecht00}. However, given the small wavelength of the 
microphases
(tens or hundreds of Angstroms), macroscopic size samples do not completely
order through 
spontaneous self assembly. Instead, oscillatory shears are commonly 
introduced in order to
accelerate long range order development over the required distances (see
Ref. \onlinecite{re:larson99} for a recent review). In practice, 
a variety of lamellar orientations are
observed depending on the architecture of the block and the parameters of the
shear
\cite{re:larson99,re:koppi92,re:patel95,re:maring97,re:leist99,re:chen98b},
while the mechanisms responsible for orientation selection are not
yet understood.  The first theoretical analysis of orientation selection
in block copolymers was conducted in the vicinity of the 
order-disorder transition of the copolymer, and addressed the effect of a {\em
  steady} shear on the growth of critical fluctuations
\cite{re:cates89}. Fluctuations along the perpendicular orientation
were shown to be less suppressed by the shear, and hence it was argued that
this orientation would be selected.
Consideration was later given to anisotropic viscosities of the
microphases, which led to different relative stabilities of uniform parallel
and
perpendicular configurations due to the different effects of thermal
fluctuations on each orientation \cite{re:goulian95}. Later work focused on
the role played by viscosity contrast between the
polymer blocks \cite{re:fredrickson94}, and showed that the perpendicular
alignment dominates for high shear rates, 
and parallel otherwise. Existing experimental phenomenology concerning
orientation selection is far more
complex than these analyses would suggest, and is seen to drastically depend
on the architecture of the block \cite{re:larson99}. The analysis that we
present  
does not rely on fluctuation effects near critical points, allows for
oscillatory shears, and explicitly incorporates hydrodynamic effects resulting
from viscosity contrast between the microphases, thus overcoming the
limitations of previous treatments. 

The experimentally relevant range of shear frequencies is well below the
inverse characteristic relaxation times of the polymer chains, and hence a
reduced description in terms of the monomer volume fraction is adopted 
\cite{re:leibler80,re:ohta86,re:fredrickson94}. According to this description,
the lamellar phase response
is solid like or elastic for perturbations directed along the lamellae normal,
and fluid like or viscous 
on the lamellar plane. In the limit of vanishing frequency, the viscous part of
the response has been assumed to be Newtonian with uniform shear viscosity,
and therefore parallel and
perpendicular orientations are degenerate and unmodified by the shear. We
address below the consequences of what we believe is the leading deviation away
from Newtonian response in the limits of low frequency and characteristic
flow scale much longer than the lamellar spacing: the
viscous stress tensor of a lamellar phase is, by reason of symmetry, the same
as that of
any other uniaxial phase (e.g., a nematic liquid crystal
\cite{re:martin72,re:degennes93}). The slowly varying local wavevector of the
lamellae
plays a role analogous to that of the director in a nematic. The rest of this
paper is devoted to the study of the effect of this assumption on the
hydrodynamic stability of the configuration shown in Fig. \ref{fig_conf}.

We assume that the dissipative part of the linear stress tensor is that of a
uniaxial, incompressible phase \cite{re:degennes93},
\begin{equation}
\sigma^{D}_{ij} = \eta D_{ij} + \alpha_1 \hat{n}_i \hat{n}_j \hat{n}_k
\hat{n}_l D_{kl} + \alpha_{56} (\hat{n}_i \hat{n}_k D_{jk} + \hat{n}_j
\hat{n}_k D_{ik})
\label{eq_Ericksen}
\end{equation}
with $i,j,k=x,y,z$, $D_{ij}=\partial_i v_j + \partial_j v_i$ with
$v_i$ the local velocity field, and $\hat{\mathbf n} = (\hat{n}_x,
\hat{n}_y, \hat{n}_z)$ denoting the {\em slowly varying}
normal to the lamellar planes. The Newtonian viscosity is $\eta$,
and $\alpha_1$ and $\alpha_{56}$ are two independent viscosity coefficients. 
The dynamic viscosity $\eta'$ ($\eta' = G''/\omega$,
with $G''$ the loss modulus and $\omega$ the shear frequency) is 
$\eta'=\eta$ for a fully
ordered perpendicular configuration ($\hat{\mathbf n}=(1,0,0)$), and
$\eta'=\eta+\alpha_{56}$ for a 
parallel orientation ($\hat{\mathbf n}=(0,0,1)$). This model is consistent with
low frequency rheology in PEP-PEE diblocks showing $\eta'_{\rm par} >
\eta'_{\rm perp}$ \cite{re:koppi92} if $\alpha_{56}>0$, and with 
PS-PI diblocks $\eta'_{\rm par} < \eta'_{\rm perp}$ \cite{re:chen98b} if
$\alpha_{56} < 0$. Given this assumption, the effective dynamic viscosity in
the two domain configuration of Fig. \ref{fig_conf} would be different in each
domain. It is known that the analogous
configuration for the case of two Newtonian fluids of different viscosity is
unstable both for steady \cite{re:yih67,re:hooper85} and
oscillatory \cite{re:king99} shears. We describe below the extension of these
results to uniaxial phases in the limit of small but nonzero Reynolds
number flows.

We consider a base state involving perpendicular (A) and parallel
(B) regions separated by a planar surface, subjected to an imposed shear
${\mathbf v}_0= \gamma \omega d \cos (\omega t) \hat{\mathbf y}$ at $z=d$, and
${\mathbf v}_0 = 0$ at $z=0$ (Fig. \ref{fig_conf}), with $\gamma$ the
shear amplitude and $d$ the system thickness. The resulting velocity field
${\mathbf v}_{A,B}=(0,V_{A,B},0)$ is the same as that of two superposed 
Newtonian fluids with viscosities $\mu_A = \eta$ and $\mu_B = \eta +
\alpha_{56}$. We then consider a small perturbation of the domain boundary as
shown in Fig. \ref{fig_conf}, and write the velocity fields in A and B as
$v_i^{A,B} = V_{A,B} \delta_{iy} + u_i^{A,B}$ ($i=x,y,z$). By expanding 
$u_i^{A,B} = \sum_{q_x,q_y} \hat{u}_i^{A,B}(q_x,q_y,z,t)
\exp [i(q_x x + q_y y)]$, substituting into the modified Navier Stokes
equation that results from the choice of Eq. (\ref{eq_Ericksen}), linearizing,
and eliminating pressure and $\hat{u}_y$, we find
\begin{eqnarray}
& Re \left [ (\partial_t + iq_y V_A)(\partial_z^2 - q^2) \hat{u}_z^{A}
 - iq_y(\partial_z^2 V_A) \hat{u}_z^{A} \right ] & \nonumber\\
& = (\partial_z^2 - q^2)^2 \hat{u}_z^A
 - \alpha_{56} q_x^2 (\partial_z^2 - q^2) \hat{u}_z^A  & \nonumber\\
& + iq_x [2\alpha_1 q_x^2 - \alpha_{56}
(\partial_z^2 - q^2)] \partial_z \hat{u}_x^A, & 
\label{eq_uz_A} \\
& Re \left [ \partial_t(\partial_z^2 - q^2) \hat{u}_x^{A} +
  iq_y(\partial_z^2 - q^2)(V_A \hat{u}_x^{A}) + \right. & \nonumber\\
& \left. 2q_xq_y(\partial_z V_A) \hat{u}_z^{A} \right ]
 = (1+\alpha_{56})(\partial_z^2 - q^2)^2 \hat{u}_x^{A} & \nonumber\\
& - 2\alpha_1 q_x^2 (\partial_z^2 - q_y^2) \hat{u}_x^{A}, &
\label{eq_ux_A}
\end{eqnarray}
for the perpendicular domain A ($0 \le z \le d_A$). Here $q^2=q_x^2 +
q_y^2$ and $Re= \rho \omega d^2 / \eta$ is the Reynolds number, with
$\rho$ the copolymer density. Similarly
\begin{eqnarray}
& Re \left [ (\partial_t + iq_y V_B)(\partial_z^2 - q^2) \hat{u}_z^{B}
 - iq_y(\partial_z^2 V_B) \hat{u}_z^{B} \right ] & \nonumber\\
& =(1+\alpha_{56})
(\partial_z^2 - q^2)^2 \hat{u}_z^{B} - 2\alpha_1 q^2 \partial_z^2 
\hat{u}_z^{B}, & \label{eq_uz_B} \\
& Re \left [ \partial_t(\partial_z^2 - q^2) \hat{u}_x^{B} +
  iq_y(\partial_z^2 - q^2)(V_B \hat{u}_x^{B}) + \right. & \nonumber\\
& \left. 2q_xq_y(\partial_z V_B) \hat{u}_z^{B} \right ]
 = (\partial_z^2 - q^2)^2 \hat{u}_x^{B} & \nonumber\\
&+\alpha_{56} (\partial_z^2 - q^2) \left ( \partial_z^2 \hat{u}_x^{B}
-iq_x \partial_z \hat{u}_z^{B} \right ) 
 -2iq_x \alpha_1 \partial_z^3 \hat{u}_z^{B}, & \label{eq_ux_B}
\end{eqnarray}
for the parallel region B ($d_A \le z \le d$). All quantities have been
made dimensionless by a length scale $d$, a
time scale $\omega^{-1}$, and rescaling viscosities by $\eta$,
i.e., $\alpha_1 \rightarrow \alpha_1 / \eta$ and 
$\alpha_{56} \rightarrow \alpha_{56} / \eta$ ($\mu_A=1$,
$\mu_B=1+\alpha_{56}$). Rigid boundary conditions are used on the planes
$z=0$ and $z=d$ ($d = 1$ after rescaling). At the interface $z=d_A +
\zeta(x,y,t)$ we have 
${\mathbf v}^{A} = {\mathbf v}^{B}$, and $(\sigma_{ij}^B-\sigma_{ij}^A)
\hat{n}_j = - \Gamma' (\partial_x^2 + \partial_y^2) \zeta \hat{n}_i
\delta_{iz}$ with $\Gamma' = \Gamma / (\eta \omega d)$ and $\Gamma$ the
interfacial tension. Also, the kinematic boundary condition for the interface is
$\left ( \partial_t + {\mathbf v}^B \cdot {\mathbf \nabla} \right )
\zeta = v_z^B$.

Equations (\ref{eq_uz_A})--(\ref{eq_ux_B}) are similar to the
Orr-Sommerfeld equation for the Newtonian case 
\cite{re:yih67,re:hooper85,re:king99}, except that $x$ and $z$ velocity fields
are now coupled (except at $q_x=0$). The solution can be found by writing 
$\hat{u}_{z,x}^{A,B} = \exp(\sigma t) \phi_{z,x}^{A,B}$ and
$\hat{\zeta} = \exp(\sigma t) h$, with $\hat{\zeta}$ the Fourier
transform of $\zeta$, and $\sigma$ the Floquet exponent yielding the
perturbation growth rate (coefficients in Eqs. (\ref{eq_uz_A})--(\ref{eq_ux_B})
proportional to $V_{A,B}$ are periodic in time).

For typical block copolymers, $\rho \sim 1$ g cm$^{-3}$, $d \sim 1$ cm, and
$\eta \sim 10^4$--$10^6$ P, resulting in $Re / \omega= 10^{-4}$--$10^{-6}$
s. Hence $Re \ll 1$ for the frequencies of interest. We further expand
the velocity, interfacial 
functions $\phi_{z,x}^{A,B}$, 
$h$, as well as the Floquet exponent $\sigma$ in powers of $Re$, and solve
Eqs. (\ref{eq_uz_A})--(\ref{eq_ux_B}) order by order. In the limit
$Re \rightarrow 0$ while keeping the surface tension
$\Gamma'$ finite, we find $\sigma = f_{z0}^B(q_x,q_y) \Gamma'$, with 
$f_{z0}^B < 0 $ for all wave numbers $q_x$ and $q_y$. Hence the interface
is stable, indicating coexistence of parallel and perpendicular
orientations in this limit.

The situation is different for small but finite values of $Re$. In this case
we need to address the order of $\Gamma'$ as well. Typical values of
$\Gamma \sim 1$ dyne/cm lead to $\Gamma' \omega = 10^{-4}$--$10^{-6}$
s$^{-1}$. Given that $Re / \omega= 10^{-4}$--$10^{-6}$ s, and that $\omega \sim
1$ s$^{-1}$ in typical experiments, we consider the distinguished limit
$\Gamma' = {\cal O}(Re)$. Writing $\Gamma'=\Gamma_1 Re$, and $\sigma =
\sigma_1 Re$, we find
\begin{equation}
\sigma_1 = f_{z0}^B(q_x,q_y) \Gamma_1  
+ \frac{1}{2} \delta^2 \gamma^2 f_{z1}^B(q_x,q_y).
\label{eq_sig1}
\end{equation}
Here $\delta = m \lambda_0$ is proportional to the viscosity contrast
$m=\mu_A/\mu_B$, with $\lambda_0 = (d_A + m d_B)^{-1}$,
$d_B=d-d_A$, and the function $f_{z1}^B(q_x,q_y)$
depends on the system parameters $\alpha_1$, $\alpha_{56}$, and on $d_A$, 
but not on the shear parameters $\gamma$ and $\omega$. Whereas $f_{z0}^B$ is
always negative, $f_{z1}^B$ can be positive so that Eq. (\ref{eq_sig1})
illustrates the
competition between the stabilizing effect of surface tension and the
destabilizing effect of the imposed shear flow. Note that $\Gamma_1$ can
be written as $\Gamma_1=1/We=\theta \omega^{-2}$, with $We$ the Weber
number and $\theta = \Gamma/(\rho d^3)$. Thus, we have
$\sigma_1=\sigma_1(\gamma^2, \omega^{-2})$, and instability is increased at
large shear amplitudes and frequencies.

\begin{figure}
\includegraphics[width=2.9in]{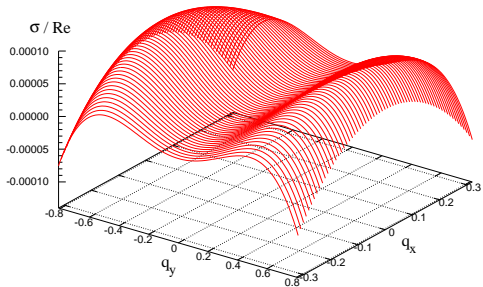}
\caption{Growth rate $\sigma /Re$ as a function of wave numbers $q_x$
  and $q_y$, for $\alpha_1=1$, $\alpha_{56}=-0.9$, $\gamma=1$,
  $Re=5 \times 10^{-4}$, and $d_A=1/2$. The maximum growth rate is found
  at $q_x^{\rm max}=0$, $q_y^{\rm max}= \pm 0.61$.}
\label{fig_sigma}
\end{figure}

Typical results for $\sigma$ as a function of wave vector are shown
in Fig. \ref{fig_sigma}. Unstable wave vectors are near $q_x = 0$. This
indicates the 
absence of interfacial modulation along the $x$ direction (the transverse
direction for perpendicular lamellae), and $u_x=0$
for the associated velocity perturbations. 
We have repeated the calculations for different ranges of parameters,
and found similar results for both $\sigma$ and velocity perturbations.

\begin{figure}
\includegraphics[width=2.7in]{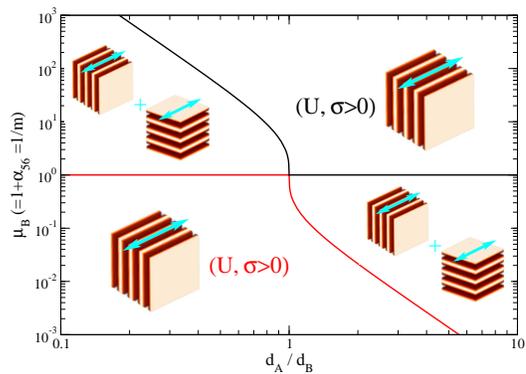}
\caption{Orientation selection for uniaxial systems under
  oscillatory shear as a function of the layer thickness ratio and viscosity
  contrast. (U, $\sigma>0$) denotes the range in which the parallel
  orientation is unstable, leading to the selection of the perpendicular
  orientation.}
\label{fig_diagram}
\end{figure}

Based on these results, further progress in determining the stability boundary
can be made by examining only long waves along the $y$ direction 
(with $q=q_y$). Equation (\ref{eq_sig1}) can be 
expanded as $f_{z0}^B = - f_0 q^4 + {\cal O}(q^6)$,
$f_{z1}^B = f_1 q^2 + f_2 q^4  + {\cal O}(q^6)$, so that the
Floquet exponent can be rewritten as
\begin{equation}
\sigma_1 = \frac{1}{2} \delta^2 \gamma^2 f_{1} q^2 - (\theta f_0 \omega^{-2} 
- \frac{1}{2} \delta^2 \gamma^2 f_{2} ) q^4,
\label{eq_sig1_q}
\end{equation}
where $f_0 > 0$ always, as noted above. Both $f_1=f_{1}(\alpha_{56}, d_{A})$
and $f_2= f_{2}(\alpha_{1},\alpha_{56},d_{A})$ are complicated but known
functions of their arguments. For small $q$ stability is determined by
the sign of $f_{1}$ which depends on $\alpha_{56}$ and layer thickness
$d_{A}$, but is independent of shear parameters.
The calculated stability diagram in the $(d_A/d_B , \mu_B)$ plane
($\mu_{B} = 1 + \alpha_{56}$) is shown in Fig. \ref{fig_diagram}. Note the
symmetry 
under $d_A/d_B \rightarrow (d_A/d_B)^{-1}$ and $\mu_B 
\rightarrow \mu_B^{-1}$ which suggests that this instability is related to the
known two fluid instability produced by viscosity stratification
\cite{re:hooper85}: instability occurs when the thinner domain is more
viscous. In the present case, however, viscosity contrast between the domains is not
caused by fluid stratification, but rather because of the effective viscosity
contrast between lamellae of different orientations. 

The stability analysis discussed thus far is purely of hydrodynamic nature,
but can we used to argue that growth of unstable modes above threshold
leads to orientation 
selection. As can be seen from Fig. \ref{fig_conf}, parallel lamellae are
marginal to velocity fields along $x$ and $y$ directions, but will be
compressed or expanded by nonuniform flows along the
$z$-direction. Conversely, perpendicular lamellae will be distorted by
nonuniform flows along $x$, but not by flows along either $y$ or $z$. Since the
instability mode is dominated by a velocity field along $z$, it will lead to
weak and oscillatory compression and expansion of the parallel region, while
leaving the perpendicular lamellae unaffected. The response of an interface
separating two lamellar phases, 
subjected to periodic expansion and compression and the other marginal, has
already been addressed in Ref. \onlinecite{re:huang03}. We showed that the
overall free energy of the system is reduced by the motion of the interface towards the
distorted phase (which is storing elastic energy during each cycle of the
shear) or, in the present case, toward the parallel region. In summary, Fig.
\ref{fig_diagram} shows the regions of parameters in
which parallel and perpendicular layers would coexist, and those in which the
layer of perpendicular orientation would grow at the expense of the parallel
layer. It would be interesting to test our predictions by examining a system comprising
only two layers of different orientations and with varying ratios
$d_{A}/d_{B}$ to directly address the stability of the configuration under
shear, and to indirectly measure the coefficient $\alpha_{56}$ from the
location of the instability threshold.

Experiments addressing orientation selection always involve coarsening of
polycrystalline samples with a distribution of grain sizes, and hence a range of ratios
$d_{A}/d_{B}$, and a distribution of orientations. It is generally argued that
lamellar domains with local wavevector not on the parallel/perpendicular
plane will be eliminated from the distribution rather quickly, and hence that the
selection of a final orientation will be determined by the competition between
parallel and perpendicular domains. Generally
speaking, our results imply selection of the perpendicular orientation
for finite shear frequencies and $\alpha_{56} > 0$, the latter case appropriate for
PEP-PEE but not PS-PI blocks if our assumption in Eq. (\ref {eq_Ericksen})
holds. If $\alpha_{56} < 0$, Fig. \ref{fig_diagram} would suggest that a
smaller than average layer of perpendicular orientation first grows at the expense
of neighboring parallel layers. Following this initial coarsening in which
$d_{A}/d_{B}$ increases in time, the stability boundary in this figure would
be reached. It is difficult to assess in this
case the impact of other dynamical factors affecting coarsening such as the
effect of an already moving boundary and the concomitant flows, or spatial
correlations built into the distribution of orientations following this
intermediate coarsening \cite{re:mullins93}. In summary, if $\alpha_{56} < 0$,
as well as in the limit of $Re \rightarrow 0$, our study indicates coexistence of
parallel and perpendicular domains, not inconsistently with experiments
addressing orientation selection that indicate dependence on processing
history or experimental details \cite{re:larson99,re:patel95,re:maring97}
(e.g., quenched or annealed history of the sample, and the starting time of
shear alignment). 

Once within the region of instability, it is instructive to analyze the
dependence of the growth rate of the most unstable perturbation on the
parameters of the shear. This growth rate is given by $\sigma_1^{\rm max}
=(\delta^4 f_1^2 / 16 f') \gamma^4 \omega^2$, with $f'=\theta f_0 -
  \delta^2 f_2 \gamma^2 \omega^2 /2$, and the corresponding most unstable
  wavenumber is $q_{\rm max}=\delta (f_1/f')^{1/2} \gamma \omega /2$, both of which
increase with shear amplitude and frequency. By noting that
$\sigma_1^{\rm max} =\sigma_{\rm max} /Re = \sigma_{\rm max} \eta 
/ (\rho d^2 \omega)$, and that usually $\sigma_{\rm max} \ll 
(\rho d^2 \delta^2 f_1^2 / 8 |f_2| \eta) \gamma^2 \omega$ for small
enough $\sigma_{\rm max}$, we find the maximum growth rate to be given by
\begin{equation}
\sigma_{\rm max} = \frac{\rho d^2 \delta^4 f_1^2}{16 \theta f_0 \eta} \gamma^4
\omega^3.
\label{eq_sig_max}
\end{equation}
Therefore, perturbation growth for a given block copolymer is constant along
the line $\gamma \omega^{3/4} = $const. To our knowledge, the only
experimental determination of the boundary in parameter space separating
regions in which parallel or perpendicular lamellae are selected has been
given for PS-PI copolymers \cite{re:maring97,re:leist99}. From a limited data
set, it was approximated by $\gamma \omega =$ const. Since it is not
inconceivable that the experimentally determined boundary does not correspond
to the true stability boundary, but rather to the line in which
$\sigma_{\rm max}$
becomes experimentally observable \cite{re:zf4}, it would be desirable to
conduct the experiment in a block copolymer with $\alpha_{56} > 0$.

In summary, by assuming that the dissipative part of the linear stress tensor of a
block copolymer has to respect the broken symmetry of uniaxial lamellar phases,
we have obtained a long wavelength hydrodynamic instability of the interface
separating lamellae of parallel and perpendicular orientations under an
imposed oscillatory shear. The instability leads to nonuniform secondary
flows, which would favor the perpendicular orientation in large regions of
parameter space. Since our results follow from the symmetry of the
microphases, we would expect them to hold in other complex
fluids of the same symmetry.

This research has been supported by the National Science Foundation
under grant DMR-0100903, and by NSERC Canada.

%\bibliographystyle{prsty}
%\bibliography{$HOME/mss/references}
\bibliography{references}

\begin{thebibliography}{10}
\bibitem{re:zf4}
An analogous situation would be the determination of the cloud point in
  nucleation studies.
\end{thebibliography}

\end{document}